\documentstyle[12pt,aasms4]{article}

\newcommand{\etal}{{\it et al. \/}}
\def\ltsima{$\; \buildrel < \over \sim \;$}
\def\gtsima{$\; \buildrel > \over \sim \;$}
\def\lsim{\lower.5ex\hbox{\ltsima}}
\def\gsim{\lower.5ex\hbox{\gtsima}}
\def\msole{~M_{\odot}}

\def\cmtre {~cm$^{3}~$}

\lefthead{Possenti \etal}
\righthead{POPULATION SYNTHESIS OF SHORT PERIOD PSRs}

\begin{document}

\title{}

\begin{center}
\begin{huge}
{\bf Population synthesis
of millisecond\\
and submillisecond pulsars} 
\end{huge}
\end{center}

\author{Andrea Possenti$^1$, Monica Colpi$^2$, Nichi D'Amico$^3$, Luciano 
Burderi$^4$}

\affil{$^1$ Dept. of Astronomy, University of Bologna, \\ Via Zamboni 33, 
40126 Bologna, Italy \\ e--mail: l\_possenti@astbo3.bo.astro.it
\\$^2$ Dept. of Physics, University of Milano, \\ Via Celoria 16, 
20133 Milano, Italy \\ e--mail: colpi@uni.mi.astro.it
\\$^3$ Astronomical Observatory of Bologna, \\ Via Zamboni 33, 
40126 Bologna, Italy \\ e--mail: damico@astbo1.bo.cnr.it
\\$^4$ Inst. of Physics, University of Palermo, \\ Via Archirafi 36, 
90123 Palermo, Italy \\ e--mail: burderi@gifpa2.fisica.unipa.it}

\begin{large}
\begin{center}
{\sl To appear in the ASTROPHYSICAL JOURNAL LETTERS}
\end{center}
\end{large}

\begin{abstract}

Known millisecond pulsars have periods longer than $1.558$ ms. 
Recycled in binary systems, neutron stars can attain very short spin periods.
In this paper we investigate the expected properties 
of the millisecond pulsar distribution by simulating
synthetic populations under different assumptions for 
the neutron star equation of state and decay of the magnetic field.
We find evidence that a tail in the distribution
of millisecond pulsars may exist at periods shorter than those observed. 

\end{abstract}

\keywords{Stars: neutron -- pulsars: general, rotation -- magnetic field.
Accretion.}

\section{Introduction}
The shortest period at which neutron stars (NSs) 
have been observed ($P_{\rm min}=1.558$ ms) 
is suspiciously close to the detection
capabilities of pulsar (PSR) surveys carried out so far. 
A new experiment at the Northern Cross
at Medicina has started recently 
whose sensitivity in the range $0.6 \to 1.5$ ms is 
significantly better than that of the previous surveys 
(D'Amico et 
al., 1998\markcite{x}). The experiment is  aimed at determining 
whether $P_{\rm min}$  is indeed a physical limit or is of 
instrumental origin.
  
The minimum period
of a stable, gravitationally bound NS 
is thought to be  within $\sim 0.6$ ms (for the softest equation of state
of dense matter) and $\sim 1.3$ ms (for the stiffest one).
These are the limits derived by Cook, Shapiro \& Teukolsky 
(1994a\markcite{cst94l}; CST hereafter)
in the context of a  recycling scenario where accretion 
occurs from the inner edge of a Keplerian disk               
onto a  bare unmagnetized  NS. 
Remarkably, submillisecond periods are attained 
(before the mass-shedding instability is encountered) 
even  in those models 
having a static maximum mass close to $1.4\msole$.

Several authors (Phinney \& Kulkarni 1994\markcite{pk94}; 
Stergioulas \& Friedman 1995\markcite{sf95}; 
Burderi \& D'Amico 
1997\markcite{bd97}) 
discussed the importance of  
detecting  an  ultra-rapid rotating
pulsar as a way of discriminating
between the proposed equations of state (EoSs).
The formation of a very fast
spinning pulsar, however, depends sensitively
on the history of the NS in the binary and on the evolution of 
the magnetic field which controls the dynamics of mass and angular momentum
transfer at the  magnetospheric boundary.
Whilst  detailed evolutionary calculations
by Urpin, Geppert \& Konenkov (1997) have shown  that
NSs move into the millisecond region  of the plane 
$\mu-P$ (magnetic moment versus period)  under rather ``ordinary'' 
conditions, the 
possibility of populating the submillisecond range  is still unknown.
Though, on theoretical grounds, bare NSs can attain periods $P<1$ ms
(CST), Nature may not provide the conditions for spinning 
a significant number of NSs to ultra short periods. 
  
In this Letter we 
investigate the efficiency of the recycling process 
in producing NSs in the yet unexplored region of the $\mu-P$ diagram
below $1.558$ ms.
To this end, we simulate the distribution of a NS population
evolved assuming three models for the magnetic field decay
and selecting two representative EoSs.

\section{Our adopted evolutionary models}

\subsection{Recycling scenario}

In our statistical analysis the NSs evolve in low mass
binaries and undergo four phases
(Illarionov \& Sunyaev 1975\markcite{is75}): {\sl ejector} $\rightarrow$
{\sl propeller} $\rightarrow$
{\sl accretor} $\rightarrow$
{\sl pulsar.}

During the ejector phase, the  stellar wind from the
companion star is blown away
by the strong radiation pressure of the young pulsar, and 
the NS spins down
only via magnetic dipole radiation 
at a rate 
$$\dot P = 3.15\times 10^{-16}~~{\mu_{26}^2}\,P^{-1}\, \rm {s~yr^{-1}},
\eqno(1)$$
where $P$ is the rotation period in seconds and $\mu_{26}$ is the
magnetic moment in units of $10^{26}$ G\cmtre.
This phase continues until the ram pressure
of the stellar wind starts 
exceeding the pulsar momentum flux at the accretion radius.
Thereafter, matter penetrates down to the magnetosphere
whose radius is  
$$R_A=9.8\times10^5~\mu_{26}^{4/7}~{\dot{m}}^{-2/7}M^{-1/7}R_{6}^{-2/7} 
\rm {cm},\eqno 
(2)$$
where $M$ is the neutron star mass in solar masses and $\dot{m}$
is the accretion rate in
units of  
${\dot {M}}_{E}=1.5\times10^{-8}~R_6\msole~$yr$^{-1}$ (Bhattacharya 
\& van den Heuvel 1991\markcite{bv91}).
In this phase, the intervening plasma and the NS can  exchange 
angular momentum. As a guideline, we assume that the NS is fed through a 
disc that develops  due to Roche--lobe overflow in the low--mass 
binary. The disc has 
an inner radius $R_{in}=\phi R_A$, 
where the dimensionless  constant $\phi$ is derived following Burderi et al.
(1998\markcite{b98}):
$$\phi^{31/28}=0.56~ \alpha^{9/35} {\dot{m}}^{3/70} M^{5/140} R_6^{3/70}
R_{A,6}^{-3/28},\eqno(3)$$
with $\alpha \simeq 1$ the Sakura-Sunyaev parameterization of the viscosity.
The rate of transfer of angular momentum $L_{tr}$ at the inner 
boundary of the disc
can be evaluated as:
$${\dot{L}}_{tr}=R_{in}^2 {\Omega}_K(R_{in}) \dot{m}~g(\omega_S ), 
\eqno(4)$$
where $\Omega_K(R_{in})$ is the Keplerian angular velocity 
at $R_{in}.$ The dimensionless torque $g,$ function of 
the fastness parameter $\omega_S=\Omega /{\Omega}_K(R_{in}),$
determines the direction of
angular momentum transfer, and  vanishes
 at a certain  critical 
value $\omega_{S,crit}$.

Denoting with  $L_{NS}=I\Omega$  the NS angular momentum (with $I$ the 
moment of inertia), we use the balance
relation
$\dot L_{NS} = \dot L_{tr}$
to derive the evolution of the angular velocity $\Omega$ of the
neutron star. When  $g$ is negative 
the NS 
magnetosphere acts as a centrifugal barrier at $R_{in}$. In this phase 
({\it propeller}), the infalling plasma is forced
into super Keplerian rotation and is swept away, 
extracting
angular momentum from the compact object which spins down.
Later in the evolution, when $g$ becomes positive ({\it accretion 
phase}) the NS spins  up. We adopt two different expressions 
for $g$, to bracket uncertainties, according to the models of  Ghosh \& 
Lamb (1991\markcite{gl}) and Wang (1996\markcite{w96}).
When mass transfer from the companion stops, the NS 
shines as a {\it pulsar}, and suffers secular spin down
by magnetic dipole
torques (eq. [1]). In this terminal phase, 
potential submillisecond pulsars migrate to the right of the
$\mu-P$ diagram, and may in principle evacuate the region of
interest to our experiment. 

\subsection{The role of the EoS}
                                 
The equation of state for nuclear matter determines the internal structure
of a NS of given  mass  and angular velocity. Both these
quantities vary during the recycling process 
in a binary system: in order 
to account for the related evolution
of $R$ and $I,$ 
we used fits to the
relativistic models calculated by Cook, Shapiro \& Teukolsky 
(1994b\markcite{cst94}).
                                
For a bare neutron star of $1.4\msole$, at the onset of 
recycling, all viable EoSs allow for terminal periods
$P\lsim 10$ ms. 
However, the minimum period attainable 
depends on the specific EoS.
We focused on two EoSs,
labeled as A and L by Arnett \& Bowers (1977\markcite{ab77}), bracketing
the set of the limiting
periods;  $0.604~$ ms for EoS A, and $1.25~$ ms for EoS L.
A is  soft, with a maximum static mass of $1.66 \msole$
and a static radius
(for a $1.4 \msole$ star) of $R_{st}=9.59~$ km. EoS L is much stiffer, with
a maximum static mass of $2.70 \msole$
and a static radius of $R_{st}=15.0~$km for the canonical mass.
Both models reach the mass-shedding limit with a modest
rotational to gravitational energy ratio. This suggests that 
nonaxisymmetric 
instabilities are of no obstacle to recycling (CST) and so we
neglect their possible influence on the evolution. 

In the accretion model of CST, the
centrifugal instability sets in when the accreted  mass reaches
about $0.3 \to 0.4 \msole.$ 
As an example, in the case of EoS A, a
$1.4 \msole$-neutron star can load $0.428 \msole$ before mass shedding
disruption, with a final gravitational mass of 
$1.77 \msole$, higher than the static limit. Such an effect enhances
the maximum attainable spin rate of a recycled pulsar.
In the simulation, we thus allow for a total
accreted mass of about  
$0.5 \msole$ for EoS A, and of 0.5 (or 1.0) $\msole$ for  EoS L
(still compatible with the stability calculations of CST). 

\subsection{Magnetic field decay scenarios}

The observations of low magnetic moments
($\mu = 3.2 \times 10^{37} (P\dot P)^{1/2}$ G\cmtre $\simeq 10^{26}\to
10^{27}$ G\cmtre)
in many binary and millisecond pulsars raises the problem  of how the surface
magnetic field can decay from the typical value of the isolated
pulsar population ($\mu \simeq 10^{29} \to 10^{31}$ G\cmtre).
An original suggestion of Taam \& van den Heuvel
(1986\markcite{tv86}) prompted the development of a series of semi-empirical
models, in which the fractional decay of the magnetic moment is related
to the amount of mass accreted by the NS (Shibazaki et al.
1989\markcite{s89}). This picture appears to be  probably too simplified
(Wijers 1997\markcite{w97}, but see 
Van den Heuvel \& Bitzaraki 1995\markcite{vb95})
and other quantities (such as $\dot {m}$) are needed
to explain  the observation. We explored three different physical
models for the surface field decay.

\subsubsection{Model KB}

The first model was proposed by Konar \& Battacharya (KB 
hereafter; 1997\markcite{kb97}),
following an idea by Geppert \& Urpin (1994\markcite{gu94}). 
In this model, the accreting plasma  
heats, initially,  the crust  reducing the time-scale for ohmic 
diffusion. As the original  crustal material is conveyed towards higher 
conductivity region, ohmic decay slows down until the field freezes 
(when the  current layers are completely assimilated into the NS 
super-conducting core).
As a result, the amount of magnetic moment decay depends not only
on the total mass accreted, but also on $\dot {m}$. In particular
higher accretion rates determine a precocious freezing of the field, which
relaxes to a higher bottom value. 
We use the same field decay profiles presented
in KB extrapolating them also to values of  $\dot {m}>0.1$
(Brown \& Bildstein 1998).
Figures 5 and 6 of KB refer to evolution for an intermediate EoS:  
As field decay depends on the properties of the crust, we 
tested the sensitivity of the results introducing 
a correction factor to the magnitude of the relic field
for EoS A (higher value) and for EoS L (lower value), to mimic
the effect of a thinner crust for A and thicker for L.
Additional corrections (e.g., thermal history of the crust)   
could play a role but are difficult to include quantitatively.
This point deserves future study.

\subsubsection{Model BLOB}
                               
In Burderi, King \& Wynn (1996\markcite{bkw96}) 
the magnetic moment decay is caused by an ordered accretion onto the 
polar caps of the compact object. Matter accreted at the inner 
radius of the disc in the form of diamagnetic plasma blobs develops 
induced currents in its impact onto the magnetic poles. 
These currents can partially
neutralize the NS crustal currents, weakening the magnetic moment
of the star. Under simplifying assumptions, it is possible to derive a 
law for the magnetic field evolution:
$$\dot{\mu} = -\kappa_{\mu} \phi^{-7/4} {\dot{m}}^{3/2} M^{1/4} R_6^{1/2}
R_{in,6}^{-1/4}, 
\eqno(5)$$
where $\kappa_{\mu} = 6.6\times 10^{-3}$.
When $R_{in}=R$ the decay stops and the related bottom magnetic moment is:
$$\mu_{bottom,26} = 1.04~\phi^{-7/4}~{\dot{m}}^{1/2}~M^{1/4}~R_{6}^{9/4}. 
\eqno(6)$$

\subsubsection{Model SIF}
                          
Another class of models involves the spin history of the NS to drive 
evolution of the field residing initially in the core  
(Srinivasan et al. 1990\markcite{s90};
Ding, Chen \& Chau 1993\markcite{dcc9}; Miri 1996\markcite{m96}).
As the NS spins down, the inter-pinning between
the proton flux tubes and the superfluid neutron vortex lines drags the
fluxoids (carrying the core field $B_{c}$) towards the crust, 
where the magnetic 
flux undergoes ohmic diffusion: the surface field $B_{s}$ then relaxes
to the value of the residual core field on a time-scale $\tau$.
In the simplest version of the model,  
when $\dot{P}>0$ the fields evolve according to:
$${\dot{B}}_{c} = - \dot{P}~P^{-1}~B_{c}; ~~{\dot{B}}_{s} = - 
(B_{s}-B_{c})/\tau, 
\eqno(7)$$
where $\tau$ is set equal to $10^8$ yr.

\section{Simulated populations}

The statistical analysis of the simulated population is
carried out using a Monte Carlo technique. Approximately
1,000 points are needed to grant stability.
The distributions of the physical parameters selected at the 
onset of evolution are described in Table 1.
We fixed an initial gravitational mass of $1.4 \msole.$
The initial spin period 
is selected from a flat probability distribution among $0.01\to 0.10$ s, and
it turned out that the simulations are quite insensitive to
the exact range of the initial values of $P$. 
To emulate    
the distribution of the magnetic moment 
we adopt the
gaussian profile from Bhattacharya et al. (1992\markcite{b92}) which, at
present, is the best representation of the observed $\mu.$

Evolutionary calculations for the low-mass binaries 
suggest an ejector phase lasting between
$10^8 \to 2\times10^9$ yr (Miri \& Bhattacharya 1994\markcite{mb94}) 
which is accompanied by negligible spontaneous decay of the magnetic field.
The propeller phase affects the final statistics
only in the SIF model for which we mimic the spin-down effect introducing a
mass transfer rate with mean value ${\dot {m}}_{prop}~\simeq 10^{-9}$. 
The efficiency factor for angular momentum extraction is chosen according
to the value suggested by Miri (1996\markcite{m96}).
During the propeller phase the orbital parameters
are unaffected by the binary evolution
and so the rate of mass propelled is kept constant.

As regard to the accretion rates ${\dot{m}}_{accr},$
during the spin up phase,  we select an interval compatible with the 
observations of LMXBs, whose typical luminosities cluster around $0.1~L_{E}$. 
In a real binary system the accretion rates are not constant.
However, during the accretion phase
the mass transfer rate reaches a steady state
after a time (Urpin \& Geppert 1996\markcite{ug96})
remarkably shorter ($t \simeq 10^{4} \to 10^{5}$ yr) than the
typical duration of this evolutionary stage in the LMXB ($t>10^{7}$ yr).
We allow for a total accreted mass $M_{accr}$ greater than $0.01 \msole$
and smaller than the maximum value permitted by the EoS (see $\S 2.2$).
The duration of the accretion phase is determined combining
${\dot{m}}_{accr}$ and $M_{accr}$, and can never exceed $2\times 10^{9}$ yr. 
To test the dependence of the results on $M_{accr}$ and indirectly on
the mass of the companion star we also considered different boundary values.
When accretion ends, the NS spins down by dipole torque for
a time determined by a flat distribution among $10^{8} \to 3\times10^9$ yr.
A longer pulsar phase does not modify the outcomes.

\section{Results and Conclusions}

The results of our simulations are summarized in Figure 1, 
where we give the fraction of synthesized objects
for EoS A and L in the 4 selected regions of the $\mu~-~P$ plane.
This fraction  is derived
normalizing the sample to the total number of pulsars
spinning with a period $P<10$ ms.
As a guideline, the upper left number in each cross
represents the relative percentage of objects
having $\mu>10^{25.6}$ G\cmtre (nowadays the minimum
value detected in the observed population of millisecond pulsars) and
a period $P<1.558$ ms (the minimum value observed).
The typical variance is about $3\%$ for the upper two quadrants of the 
cross.

We find that {\it NSs with periods 
below $1.558$} ms {\it are present 
in a statistically significant number}, both in KB and BLOB models.
As regard to the sensitivity of the results on the adopted parameters, 
we did not encounter major statistical differences
when varying the  torque function $g.$ 
The percentage of ultra rapid spinning PSRs is  strongly
depressed only when the relic magnetic field is an order
of magnitude greater than that calculated in Konar \& Bhattacharya (1997).

As an illustration of the results of our runs, Figure 2 plots
$\mu$ versus $P$  for KB (top), BLOB and SIF (for EoS A). 
The first two scenarios give quite similar distributions
below  $P< 0.01$ s, whilst in SIF model, the 
population cluster between $0.01 \to 1.00$ s, leaving the
region below $1.558$ ms devoid of objects.
In the BLOB model, NSs with relatively high $\mu$ and long period
populate the plane and this is a consequence of equation (6)
predicting a bottom field higher than that found in KB, for
the same $\dot {m}$.

In summary we find that:  

(a) The efficiency of the recycling process in spinning the NSs
to the ultrashort periods can reach values as high as 
$20\%$.
{\it This is the first indication that a tail in the distribution of
the millisecond NSs may exist, at periods shorter than 
those effectively searched so far}.
Our results provide an estimate of the relative importance
of such a tail with respect to the bulk of the known millisecond PSRs. 

(b) As regard to the issue on the EoS,  
only in the BLOB scenario the distributions of 
ultra rapidly spinning PSRs enable us to discriminate between the EoS A and L.
In the KB model, they are almost indistinguishable.
A more pronounced difference in the distributions is found 
when we limit $M_{accr}$ to 0.5$\msole$ for EoS L 
($4\% \pm 1\%$ for the left  upper quadrant of KB model
to contrast with the estimate of $9\%\pm3\%$ for $M_{accr}=1\msole$).
   
(c) In the millisecond range, the distribution of $\rm {Log} \mu$   
privileges values $> 25.6$.
This comes from the hypothesis that the present millisecond pulsars 
evolved from the Low Mass X-Ray Binary population, with typical
observed accretion rates ${\dot{m}}_{accr}$ clustering around $0.1.$ 
If future observations will be in favor of low values of $\mu$, 
it will be suggestive of the presence of a yet  undetected 
population of low luminosity low mass binaries. 

(d) Pulsar search experiments with a sensitivity profile almost flat
in the ultrashort period range, like the new one in progress at the
Northern Cross radiotelescope (D'Amico et al. 1997), are strongly
encouraged, as they might help putting constraints on the equation
of state and on the evolutionary paths of recycled neutron stars. 

\acknowledgments
We thank D. Bhattacharya, U. Geppert and V. Urpin  
for helpful discussions and the Referee D. Lorimer for
enlighting comments and a critical reading of the manuscript.

\newpage

\figcaption[apjevolnew_fig1.ps]{Fraction of synthesized objects for EoS A and 
EoS L in 4 selected regions of the Log$P$-Log$\mu$ plane.\label{fig1}} 

\figcaption[apjevolnew_fig2.ps]{Distributions obtained for EoS A with model
KB (top), BLOB and SIF (bottom). The position of PSR B1937+21 
is denoted with a magnified dot. 
\label{fig2}}

\newpage


\begin{deluxetable}{cccc}
\tablecolumns{4}
\tablewidth{0pt}
\tablecaption{Starting values and evolution parameters
for population syntheses\label{tab1}} 
\tablenum{1}
\tablehead{
\colhead{Physical quantity} &
\colhead{Distribution} &
\colhead{Values} &
\colhead{Units} 
}
\startdata 
 Initial NS Mass        &  One-value    &                        1.40                                &  $M_{\odot}$
\\
 Initial NS Period      &     Flat      &                  0.01 $~~\to~~$ 0.10                       &   sec
\\
 Initial NS $~\mu$      &   Gaussian    &  Log($\mu_{ave}$)$~$ = $~$30.36$~~~~~~~~\sigma$=0.32       &  G~cm$^{3}$
\\
  ${\dot{m}}_{prop}$    &   Gaussian    &  Log(${\dot{m}}_{prop,ave}$) = --9.0$~~~~~\sigma$=1.00     & ${\dot M}_{E}$
\\
  ${\dot{m}}_{accr}$    &   Gaussian    &  Log(${\dot{m}}_{accr,ave}$) = --1.0$~~~~~\sigma$=0.50     & ${\dot M}_{E}$
\\
 Minimum accreted Mass  &   One-value   &                       0.01                                 & $M_{\odot}$
\\
  Ejector phase time    &    Flat       &          10$^{8}~~\to~~$2$~\times~$10$^{9}$                &    year
\\
 Accretion phase time   &  Flat in Log  &          10$^{6}~~\to~~$3$~\times~$10$^{9}$                &    year
\\
 Propeller phase time   &    Flat       &          10$^{8}~~\to~~$2$~\times~$10$^{9}$                &    year
\\
  Pulsar phase time     &    Flat       &          10$^{8}~~\to~~$3$~\times~$10$^{9}$                &    year
\\
\enddata
\end{deluxetable}

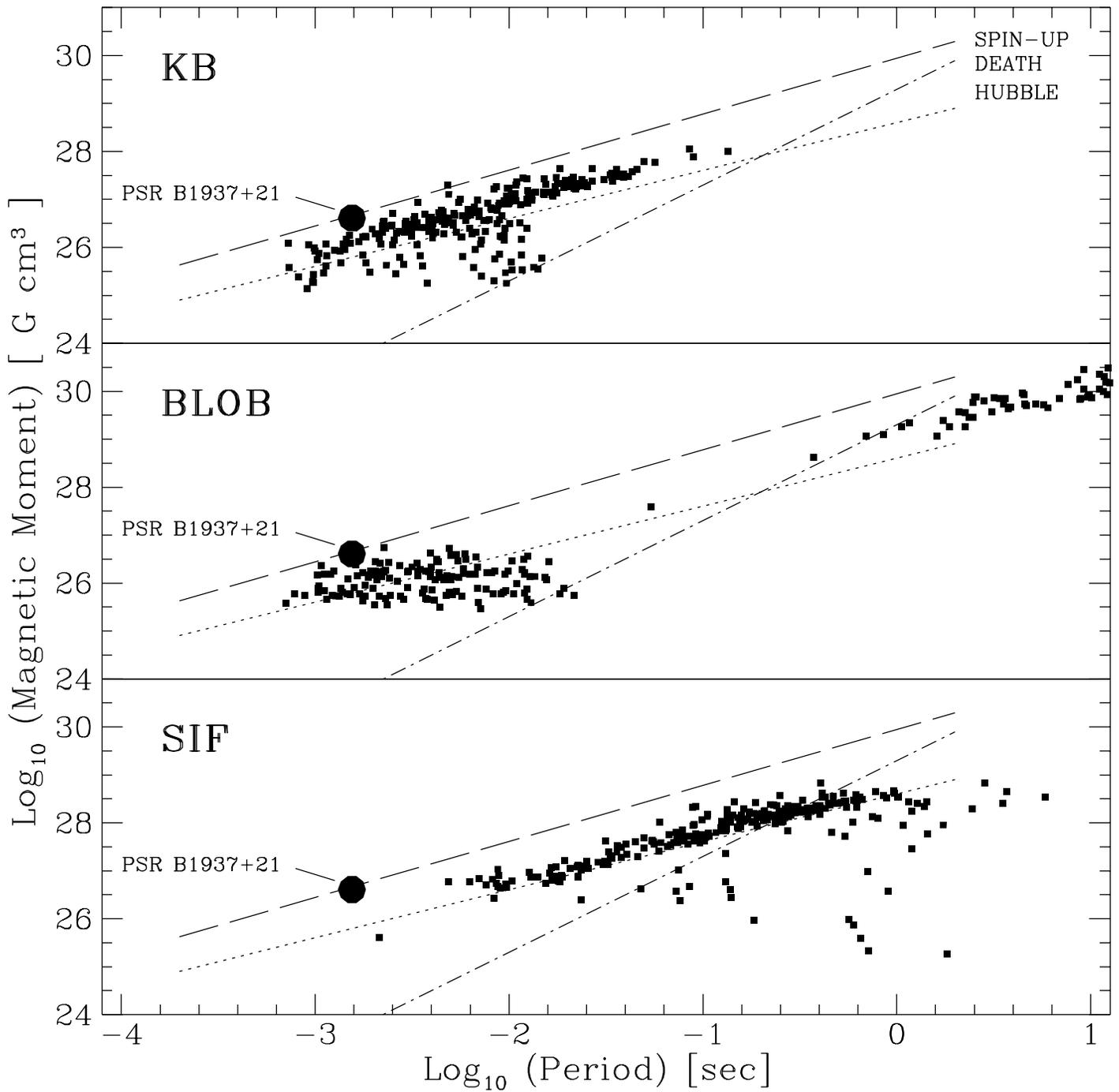
\end{document}